\documentclass{svjour2}                    
\smartqed  
\usepackage{graphicx}
\begin{document}

\title{Phase transition of triangulated spherical surfaces with elastic skeletons
}


\author{Hiroshi Koibuchi}


\institute{H. Koibuchi \at
Department of Mechanical and Systems Engineering, Ibaraki National College of Technology, Nakane 866, Hitachinaka, Ibaraki 312-8508, Japan
 \\
              \email{koibuchi@mech.ibaraki-ct.ac.jp}           
}

\date{Received: date / Accepted: date}

\maketitle

\begin{abstract}
A first-order transition is numerically found in a spherical surface model with skeletons, which are linked to each other at junctions. The shape of the triangulated surfaces is maintained by skeletons, which have a one-dimensional bending elasticity characterized by the bending rigidity $b$, and the surfaces have no two-dimensional bending elasticity except at the junctions. The surfaces swell and become spherical at large $b$ and collapse and crumple at small $b$. These two phases are separated from each other by the first-order transition. Although both of the surfaces and the skeleton are allowed to self-intersect and, hence, phantom, our results indicate a possible phase transition in biological or artificial membranes whose shape is maintained by cytoskeletons.
\end{abstract}

\section{Introduction}\label{intro}
The so-called hop-diffusion of membrane proteins or lipids observed in cell membranes indicates that the cytoskeleton forms compartments on the surface \cite{Kusumi-BioJ-2004}. It has also been recognized that the cytoskeleton of cell membranes maintains its cell shape against external forces, and plays important roles in cell motion, deformation and in some other functions. The mechanical strength of the cell is considered to be provided by cytoskeletons. Some artificial membranes are considered to have skeletons, because they are partly polymerized \cite{CNE-PRL-2006}.

A well-known conventional model for such cell membranes is the curvature model of Helfrich, Polyakov and Kleinert (HPK) \cite{HELFRICH-1973,POLYAKOV-NPB1986,KLEINERT-PLB1986}. The cell membranes are considered to be a two-dimensional surface in the HPK prescription, and therefore the two dimensional differential geometry is suited to describe the shape and their corresponding properties \cite{NELSON-SMMS2004,Gompper-Schick-PTC-1994,Bowick-PREP2001}. However, the skeleton is not considered in those curvature models.

 Skeleton models for the cytoskeleton were investigated in \cite{BBD-BioPJ-1998}. A hard-wall and a hard-core potential were assumed on the polymer chains with junctions, and the responses to some external stresses and the compression modulus were extracted from the Monte Carlo (MC) simulation data \cite{BBD-BioPJ-1998}. The compartmentalized structure was recently investigated in the framework of the HPK model \cite{KOIB-COND-MAT-0607224}. A fluid surface undergoes a phase transition due to the compartment.

However, phase transitions of skeleton surface models are yet to be studied. The terminology {\it skeleton surface model} in this paper denotes a surface model with skeletons; that is {\it surface} + {\it skeleton}, where the {\it surface} denotes the ordinary two-dimensional surface and the {\it skeleton} denotes one-dimensional elastic chains with junctions. It is expected that skeletons can make the surface smooth against the surface fluctuations. Therefore it is interesting to see whether the crumpling transition occurs in such skeleton surface models. The transition is the one that has long been studied theoretically \cite{Peliti-Leibler-PRL1985,David-Guitter-EPL1988,PKN-PRL1988} and numerically \cite{KANTOR-NELSON-PRA1987,KD-PRE2002,KOIB-PRE-2005,KOIB-NPB-2006,Baum-Ho-PRA1990,CATTERALL-NPBSUP1991,AMBJORN-NPB1993,KOIB-EPJB-2005} on the basis of the HPK model, and an experimental investigation on the transition has also been performed recently \cite{CNE-PRL-2006}.

We consider that the problem of whether the skeleton surface model undergoes the phase transition or not is non-trivial. In fact, we know that one-dimensional objects, such as an elastic ring obeying a Hamiltonian for the local interactions, has no phase transition.  
 
In this paper, we study the phase structure of a spherical skeleton surface model, which is defined on a triangulated spherical surface with polymer chains linked to each other at junctions, focusing on how the smooth phase separated from the crumpled phase. The Hamiltonian is a linear combination of the Gaussian bond potential $S_1$, the one-dimensional bending energy $S_2$ on the polymer chain, and the two-dimensional bending energy $S_J$ at the junctions. 

The interaction in the model is not purely one-dimensional. One-dimensional polymer chains have two-dimensional interactions through the two-dimensional potential $S_1$ and the two-dimensional bending energy $S_J$ at the junctions, which are considered to be the boundary of polymer chains. However, it is obvious that the elastic skeletons maintain the shape of the membrane, because triangulated surfaces have no two-dimensional bending elasticity except at the junctions of skeletons.  

We will find a first-order transition in the model under certain conditions on the length of the polymer chains and the bending rigidity for $S_J$. Although the model is allowed to self-intersect and, hence, phantom, our results indicate a possible phase transition in biological or artificial membranes whose mechanical strength is provided by the cytoskeletons.

\section{Model}\label{model}
\begin{figure}[htb]
\unitlength 0.1in
\begin{picture}( 0,0)(  10,10)
\put(16,8.5){\makebox(0,0){(a) $(2562,600,42,6)$ }}%
\put(33,8.5){\makebox(0,0){(b) $(4002,840,42,8)$ }}%
\end{picture}%
\vspace{0.5cm}
\centering
\includegraphics[width=8.5cm]{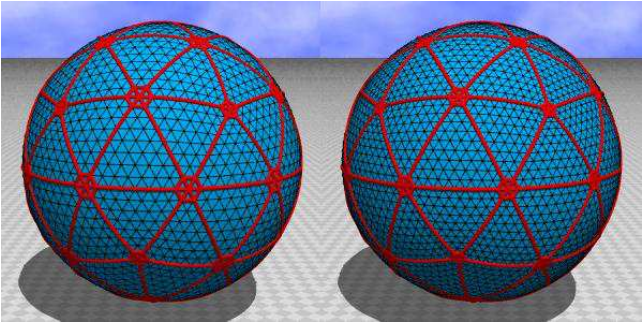}
\caption{Starting configurations of (a) $(N,N_S,N_J,L)\!=\!(2562,600,42,6)$ and  (b) $(N,N_S,N_J,L)\!=\!(4002,840,42,8)$, where $N$ is the total number of vertices, $N_S$ is the total number of vertices on the chains, $N_J$ is the total number of junctions, and $L$ is the length of chains between junctions. Thick lines denote the compartment boundary, which forms chains and junctions.} 
\label{fig-1}
\end{figure}

The triangulated surfaces are characterized by $N$ the total number of vertices, $N_S$ the total number of vertices on the chains, $N_J$ the total number of junctions, and $L$ the length of chains between junctions. Figures \ref{fig-1}(a) and \ref{fig-1}(b) show surfaces of size  $(N,N_S,N_J,L)\!=\!(2562,600,42,6)$ and $(N,N_S,N_J,L)\!=\!(4002,840,42,8)$, respectively. They are assumed to be the starting configurations of the MC simulations. Thick lines denote the compartment boundary, which forms chains and the junctions. All the vertices can fluctuate only locally, and they are prohibited from diffusing because of the fixed connectivity nature of the lattice. 

A triangulated surface of size $N\!=\!10\ell^2\!+\!2$ (= the total number of vertices) is obtained by dividing every edge of the icosahedron into $\ell$ pieces of uniform length; $\ell$ is the edge length of triangles in the icosahedron. The configurations are thus characterized by $N_5\!=\!12$ and $N_6\!=\!N\!-\!12$, where $N_q$ is the total number of vertices with co-ordination number $q$. Compartment structures are obtained by dividing $\ell$ further into $m$ pieces ($m\!=\!1,2,\cdots$), and we have chains of the uniform length $L\!=\!(\ell /m) \!-\!2$. The reason for the subtraction of $-2$ is because of the junctions at the two end points of the chain. On the surfaces in Figs. \ref{fig-1}(a) and \ref{fig-1}(b), we have $L\!=\!6$, $\ell\!=\!16$ and $L\!=\!8$, $\ell\!=\!20$, respectively.

The compartment structures shown in Figs. \ref{fig-1}(a) and \ref{fig-1}(b) are identical to those from the model in Ref. \cite{KOIB-COND-MAT-0607224}, where the boundary bonds of the compartment were kept unflipped in the dynamical triangulation. The boundary of compartment in Ref. \cite{KOIB-COND-MAT-0607224} is identical with the linear chains with junctions in this paper. The junctions form hexagons or pentagons; the total number of pentagon is $12$ and the remaining junctions are hexagons. The total number of the compartments depends on the surface size $N$, and therefore, it is increased with increasing $N$. However, the chain length $L$ can be chosen to be constant and independent of $N$. We fix the chain length $L$ to the following two values:
\begin{equation}
\label{number-inside}
L=6,\quad L=8,
\end{equation}
 which respectively correspond to the values $n\!=\!21$,  $n\!=\!36$, the total number of vertices inside a compartment \cite{KOIB-COND-MAT-0607224}. The reason why we fix $n$ is that the size of compartment is considered to be finite and, thus, it is expected that total number of lipids in the compartment also remains finite in the cell membranes. Note also that $n$ does not exactly correspond to the total number of lipids in a compartment of cell membranes.

The Hamiltonian of the model is given by a linear combination of the Gaussian bond potential $S_1$, the one-dimensional bending energy $S_2$, and the two-dimensional bending energy $S_J$, which are defined by
\begin{eqnarray}
\label{Disc-Eneg} 
S_1=\sum_{(ij)} \left(X_i-X_j\right)^2,\quad S_2=\sum_{(ij)} (1-\cos \theta_{(ij)}), \nonumber \\
S_J=\sum_{\langle ij \rangle} (1-{\bf n}_i \cdot {\bf n}_j),
\end{eqnarray} 
where $\sum_{(ij)}$ in $S_1$ is the sum over the bond $(ij)$ connecting the vertices $i$ and $j$, and $\sum_{(ij)}$ in $S_2$ is the sum over bonds $i$ and $j$, which contain not only bonds in the chains but also bonds that connect the center of the junction and the neighboring vertices of the chains. $S_1$ is defined not only on the skeleton but also on the surface, while $S_2$ is defined only on the chains. $\sum_{\langle ij \rangle}$ in $S_J$ is the sum over  triangles $i$ and $j$, which share the central point of the junction as the common vertex and form a hexagonal or pentagonal junction. 
The symbol $\theta_{(ij)}$ in $S_2$ is the angle between the bonds $i$ and $j$, and ${\bf n}_i$ in  $S_J$ is the unit normal vector of the triangle $i$ at the junctions.

\begin{figure}[htb]
\centering
\includegraphics[width=8.5cm]{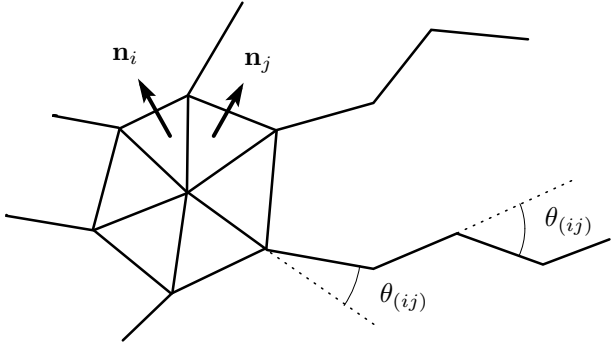}
\caption{A hexagonal junction linked to chains. The unit normal vector ${\bf n}_i$ in $S_J$ is defined on the triangle $i$ in the hexagon, and the angle $\theta_{(ij)}$ in $S_2$ is defined not only on the vertices of chains but also on those nearest to the center of junction. } 
\label{fig-2}
\end{figure}

Figure \ref{fig-2} shows a hexagonal junction linked to chains, the unit normal vectors ${\bf n}_i$, ${\bf n}_j$, and the angle $\theta_{(ij)}$, which is defined not only on the vertices of chains but also on those nearest to the center of the junction. 

We note that $S_1$ includes bonds which are the boundary edges of the hexagons (or the pentagons). The reason of this inclusion is for the sake of an in-plane elasticity of surfaces at the junctions. If these bonds were excluded from $S_1$, chains would be able to freely move into the in-plane directions at the junctions.

The partition function $Z$ of the model is defined by
\begin{eqnarray} 
\label{Part-Func}
 Z = \int^\prime \prod _{i=1}^{N} d X_i \exp\left[-S(X)\right],\\  
 S(X)=S_1 + b S_2 + b_J S_J, \nonumber
\end{eqnarray} 
where $b$ is the one-dimensional bending rigidity, $b_J$ is the two-dimensional bending rigidity at the junctions, and  $\int^\prime$ denotes that the center of the surface is fixed. In this paper, $b_J$ is fixed to
\begin{equation}
\label{two-dim-rigidity}
b_J=10,
\end{equation}
so that the junctions are sufficiently smooth. The value $b_J\!=\!10$ is relatively larger than the first-order transition point $b_c\!\simeq\! 0.8$ in the tethered surface model \cite{KOIB-PRE-2005,KOIB-NPB-2006}. Therefore, the hexagonal or pentagonal junctions are almost flat even when the surface is in the crumpled phase at sufficiently small $b$.

Both $b$ and $b_J$ have units of $kT$, where $k$ is the Boltzmann constant, and $T$ is the temperature. The surface tension coefficient $a$ of $S_1$ is fixed to $a\!=\!1$; this is always possible because of the scale invariant property of the model. In fact, from the expression $aS_1 \!+\! b S_2 \!+\! b_J S_J$ we immediately understand that $a\!=\!1$ is possible, because the factor $a$ of $S_1$ can be eliminated due to the scale invariance of the partition function. Since the unit of $a$ is $(1/{\rm length})^2$, the length unit of the model is given by $\sqrt{1/a}$. We use the unit of length provided by $\sqrt{1/a}\!=\!1$ in this paper, although $a$ can arbitrarily be chosen.  

The difference between the model in this paper and the conventional tethered surface model such as the one in \cite{KOIB-PRE-2005} must be emphasized. The surface shape is maintained by the two-dimensional bending energy defined all over the surface in the conventional surface model. On the contrary, the surface shape of the model in this paper is maintained by the one-dimensional $S_2$ and two-dimensional $S_J$; both are defined only on the skeletons.  

\section{Monte Carlo technique}\label{MC-Techniques}
The canonical Metropolis technique is used to update the variable $X$. The update of $X$ is divided into two steps. The first is the update of $X$ in the chains: the $X$ are shifted so that $X^\prime \!=\! X\!+\!\delta X$, where $\delta X$ is randomly chosen in a small sphere. The new position $X^\prime$ is accepted with the probability ${\rm Min}[1,\exp(-\Delta S)]$, where $\Delta S\!=\! S({\rm new})\!-\!S({\rm old})$. The second step involves the update of $X$ in the junctions, and this is further divided into three processes: the first is a random and simultaneous shift of $7$ (or $6$) vertices of the hexagon (or the pentagon) including the central vertex, and the second and the third are a random translation and a random rotation of these vertices by assuming the hexagon (or the pentagon) to be a rigid object. All of these MC processes are performed under about a $50\%$ acceptance rate, which is controlled by small numbers fixed at the beginning of the simulations.  We introduce the lower bound of $1\times 10^{-8}$ for the area of triangles. No lower bound is imposed on the bond length. 

We use surfaces of size $(N,N_S,N_J,L)\!=\!(2562,600,42,6)$, $(5762,1350,92,6)$, $(10242,2400,162,6)$, and $(16002,3750,252,6)$  for the length $L\!=\!6$,  and $(N,N_S,N_J,L)\!=\!(4002,840,42,8)$, $(9002,1890,92,8)$, and $(16002,3360,162,8)$ for the length $L\!=\!8$. A random number sequence called Mersenne Twister \cite{Matsumoto-Nishimura-1998} is used in the simulations.
\section{Results}\label{Results}
\begin{figure}[htb]
\centering
\vspace{0.7cm}
\unitlength 0.1in
\begin{picture}( 0,0)(  10,10)
\put(18.,43.5){\makebox(0,0){(a) Smooth surface at $b\!=\!8.1$ }}%
\put(35,43.5){\makebox(0,0){(b)  Collapsed surface at $b\!=\!8$}}%
\put(16,8.5){\makebox(0,0){(c) The surface section }}%
\put(33.5,8.5){\makebox(0,0){(d) The surface section}}%
\end{picture}%
\vspace{0.5cm}
\includegraphics[width=8.5cm]{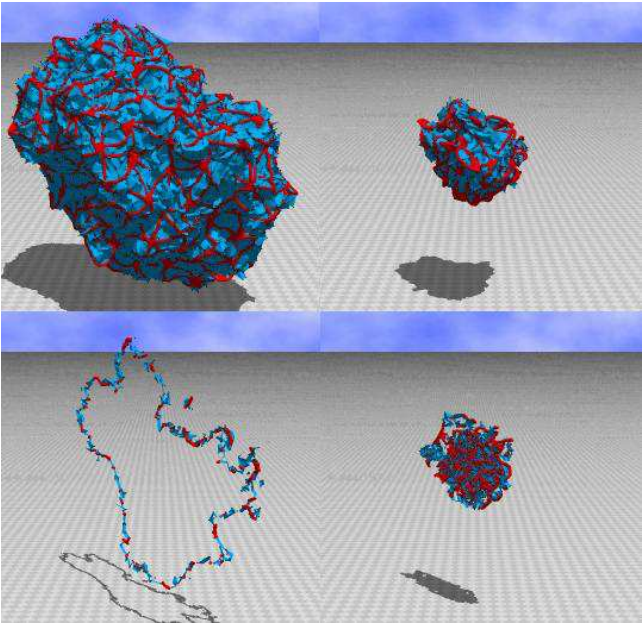}
\caption{Snapshot of the surface of size $(N,N_S,N_J,L)\!=\!(16002,3750,252,6)$ obtained in the smooth phase at (a) $b\!=\!8.1$ and in the crumpled phase at (b) $b\!=\!8$, both of which are close to the transition point. The mean square sizes of $X^2$ are (a) $X^2\!\simeq\!101$  and (b) $X^2\!\simeq\!10.1$. } 
\label{fig-3}
\end{figure}
First, we show snapshots of surface obtained at $b\!=\!8.1$ (smooth phase) and $b\!=\!8$ (crumpled phase) in Figs.\ref{fig-3}(a) and \ref{fig-3}(b), respectively. The size of surface is $(N,N_S,N_J,L)\!=\!(16002,3750,252,6)$. The surface sections are shown in  Figs.\ref{fig-3}(c),\ref{fig-3}(d). We immediately find that the surface is smooth (crumpled) at $b\!=\!8.1$ ($b\!=\!8$). 

The surface size is reflected in the mean square size of $X^2$, which is defined by
\begin{equation}
\label{X2}
X^2={1\over N} \sum_i \left(X_i-\bar X\right)^2, \quad \bar X={1\over N} \sum_i X_i,
\end{equation}
where $\bar X$ is the center of the surface. We have $X^2\!\simeq\!101$ and $X^2\!\simeq\!10.1$ on the surfaces in Figs. \ref{fig-3}(a) and \ref{fig-3}(b), respectively.

\begin{figure}[htb]
\centering
\includegraphics[width=10cm]{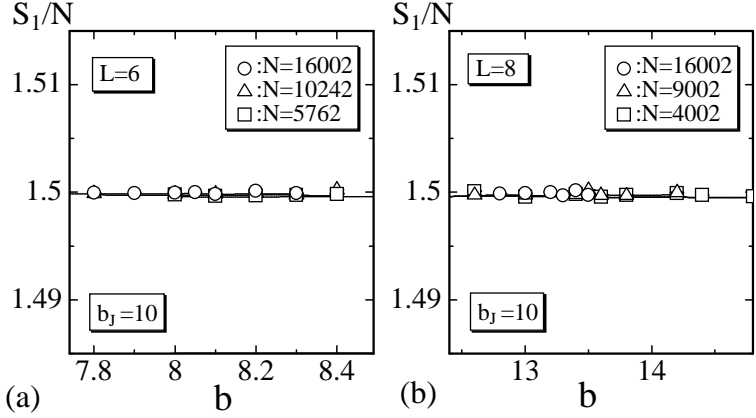}
\caption{The Gaussian bond potential $S_1/N$ vs. $b$ obtained on the surfaces of (a) $L\!=\!6$ and (b) $L\!=\!8$.  $S_1/N$ is predicted to be $S_1/N\!\simeq\!1.5$. The two-dimensional bending rigidity $b_J$ at the junctions was fixed to $b_J\!=\!10$.} 
\label{fig-4}
\end{figure}

The Gaussian bond potential $S_1/N$ against $b$ is shown in Figs. \ref{fig-4}(a) and \ref{fig-4}(b), where the length $L$ of chains between junctions are $L\!=\!6$ and $L\!=\!8$, respectively. The surface size is $N\!=\!5762$,  $N\!=\!10242$, and $N\!=\!16002$ for $L\!=\!6$, and $N\!=\!4002$,  $N\!=\!9002$, and $N\!=\!16002$ for $L\!=\!8$.  The solid lines connecting the data are drawn by a multihistogram reweighting technique \cite{Janke-histogram-2002}. Because of the scale invariant property of the partition function, we have $S_1/N\!=\!3(N-1)/2N\!\simeq\!1.5$. All of the results presented in Figs. \ref{fig-4}(a),\ref{fig-4}(b) satisfy the expected relation. 

\begin{figure}[htb]
\centering
\includegraphics[width=10cm]{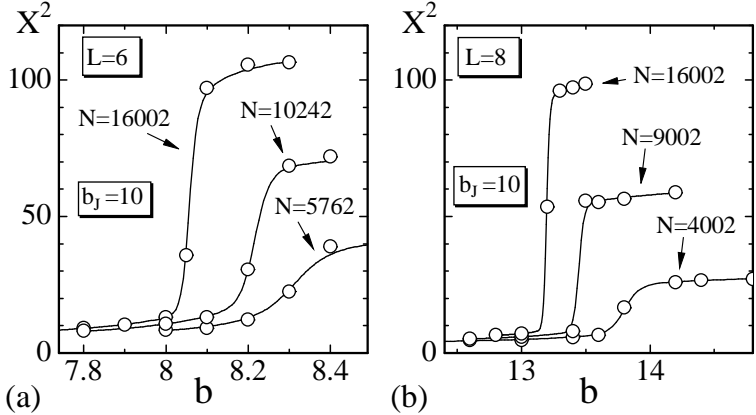}
\caption{The mean square size of $X^2$ versus $b$ is obtained on the surfaces of (a) $L\!=\!6$ and (b) $L\!=\!8$. The rigidity $b_J$ was fixed to $b_J\!=\!10$. The curves are drawn by the multihistogram reweighting technique.   } 
\label{fig-5}
\end{figure}

The mean square size of $X^2$ versus $b$ is plotted in Figs. \ref{fig-5}(a) and \ref{fig-5}(b),
where $L\!=\!6$ and $L\!=\!8$, respectively. The data was obtained on the same sized surfaces as those in  Figs. \ref{fig-4}(a) and \ref{fig-4}(b). The solid lines drawn on the data were obtained by a multihistogram reweighting technique. We find a sharp change of $X^2$ both in Figs. \ref{fig-5}(a) and \ref{fig-5}(b) as $N$ increases. This sharp change of $X^2$ seems discontinuous and hence indicates a first-order transition. We find also that the transition point $b_c$ moves left on the $b$-axis as $N$ increases in both cases (i.e., $L\!=\!6$ and $L\!=\!8$). This implies that the finite-size effect still remained on large surfaces of size $N\!=\!16002$. However, we should recall that the bending energy $S_2$ is defined on the chains of size $N_S\!=\!3750$ and $N_S\!=\!3360$, which are relatively small, on such large surfaces. It is expected that more and more large surfaces are necessary to remove the finite-size effect. We find also that the surface softens as the length of chain increases. In fact, the transition point $b_c$ moves right on the $b$-axis as $L$ increases from $L\!=\!6$  to $L\!=\!8$, as can be seen in Figs. \ref{fig-5}(a) and \ref{fig-5}(b).     

It must be noted that a large surface in the collapsed phase hardly turns into the smooth phase even at the transition point. This phenomenon seems typical to surface simulations \cite{KOIB-PRE-2005,KOIB-NPB-2006} based on the canonical MC simulation technique. When a large surface configuration is once trapped in a potential minimum, the configuration appears almost confined inside the potential valley. These troublesome phenomena can be avoided with more sophisticated MC techniques \cite{Berg-Neuhaus-PRL1992,Berg-Celik-PRL1992}. 

We also comment on the Hausdorff dimension $H$, which is defined by the scaling relation $X^2 \propto N^{2/H}$ and was not calculated explicitly.  However, it is obvious that $H\!\simeq\! 2$ is satisfied in the smooth phase close to the transition point, because the spherical surfaces are swollen and smooth at that point as can be seen in the snapshot of Figs. \ref{fig-3}(a) and \ref{fig-3}(c). Moreover, we expect that $H$ is greater than the physical bound $H>3$ in the crumpled phase close to the transition point. In fact, we find from Figs. \ref{fig-5}(a) and \ref{fig-5}(b) that $X^2$ appears to remain unchanged as $N$ increases in the crumpled phase. This implies that $H$ is very large and is in contrast to that of the conventional tethered surface model in \cite{KOIB-PRE-2005,KOIB-NPB-2006}, where $H$ is less than the physical bound in the crumpled phase.

\begin{figure}[htb]
\centering
\includegraphics[width=10cm]{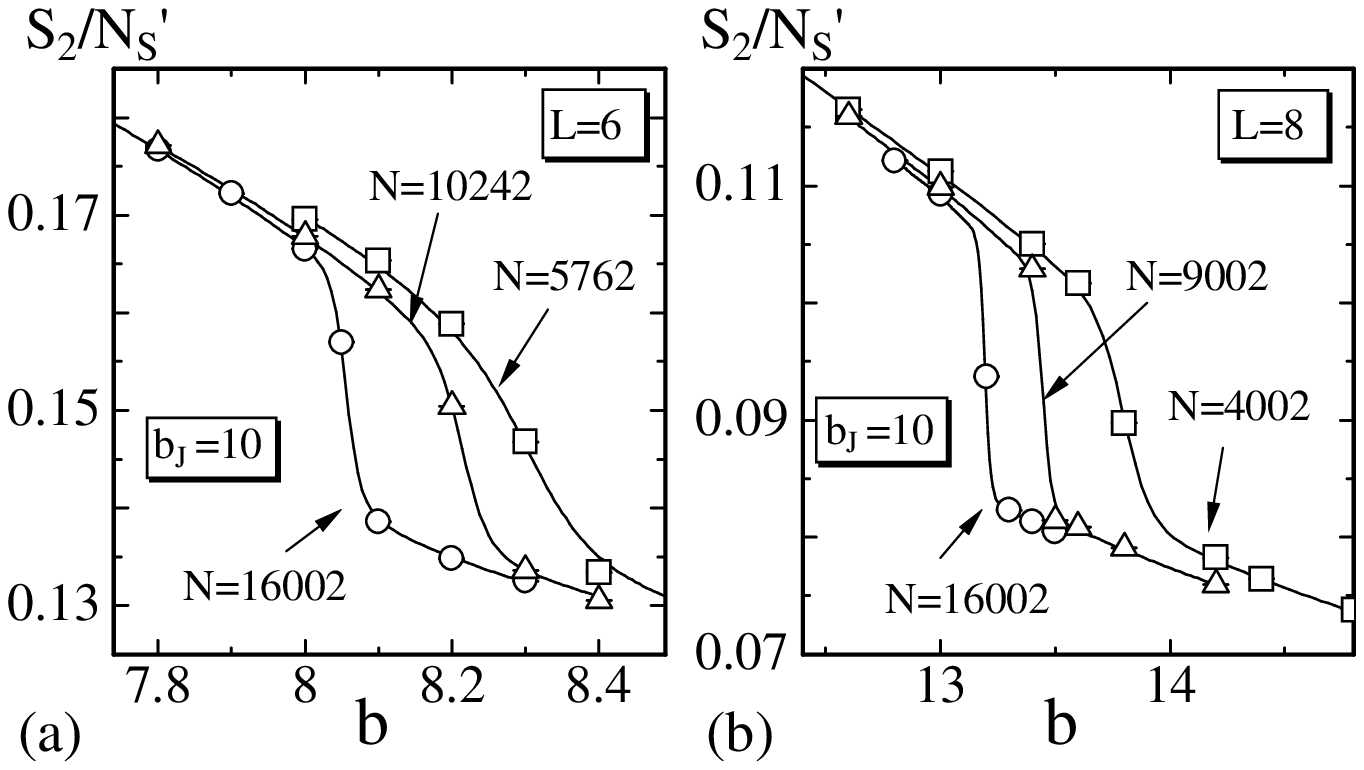}
\caption{The one-dimensional bending energy $S_2/N_S^\prime$ vs. $b$ obtained on the surfaces of (a) $L\!=\!6$ and (b) $L\!=\!8$. } 
\label{fig-6}
\end{figure}

The bending energy $S_2/N_S^\prime$ obtained on the surfaces of $L\!=\!6$ and $L\!=\!8$ is shown in Figs. \ref{fig-6}(a) and \ref{fig-6}(b), where $N_S^\prime\!=\!N_S+6N_J-12$ is the total number of vertices on which $S_2$ is defined. We have $N_S^\prime\!=\!5250$ for the surface of $(N,N_S,N_J,L)\!=\!(16002,3750,252,6)$ and $N_S^\prime\!=\!4320$ for that of $(N,N_S,N_J,L)\!=\!(16002,3360,162,8)$. Although the data in the figures are connected with smooth curves obtained by a multihistogram reweighting technique, the discontinuous nature of $S_2/N_S^\prime$ is almost apparent. 

The phase transition is not reflected in the two-dimensional bending energy $S_J$.  Since the corresponding bending rigidity $b_J$ was chosen to make $b_J\!=\!10$ (as shown in Eq.(\ref{two-dim-rigidity})), the junctions are maintained smooth enough. Therefore, $S_J$ defined on the junctions is kept small and exhibits only weak fluctuations.

\begin{figure}[htb]
\centering
\includegraphics[width=10cm]{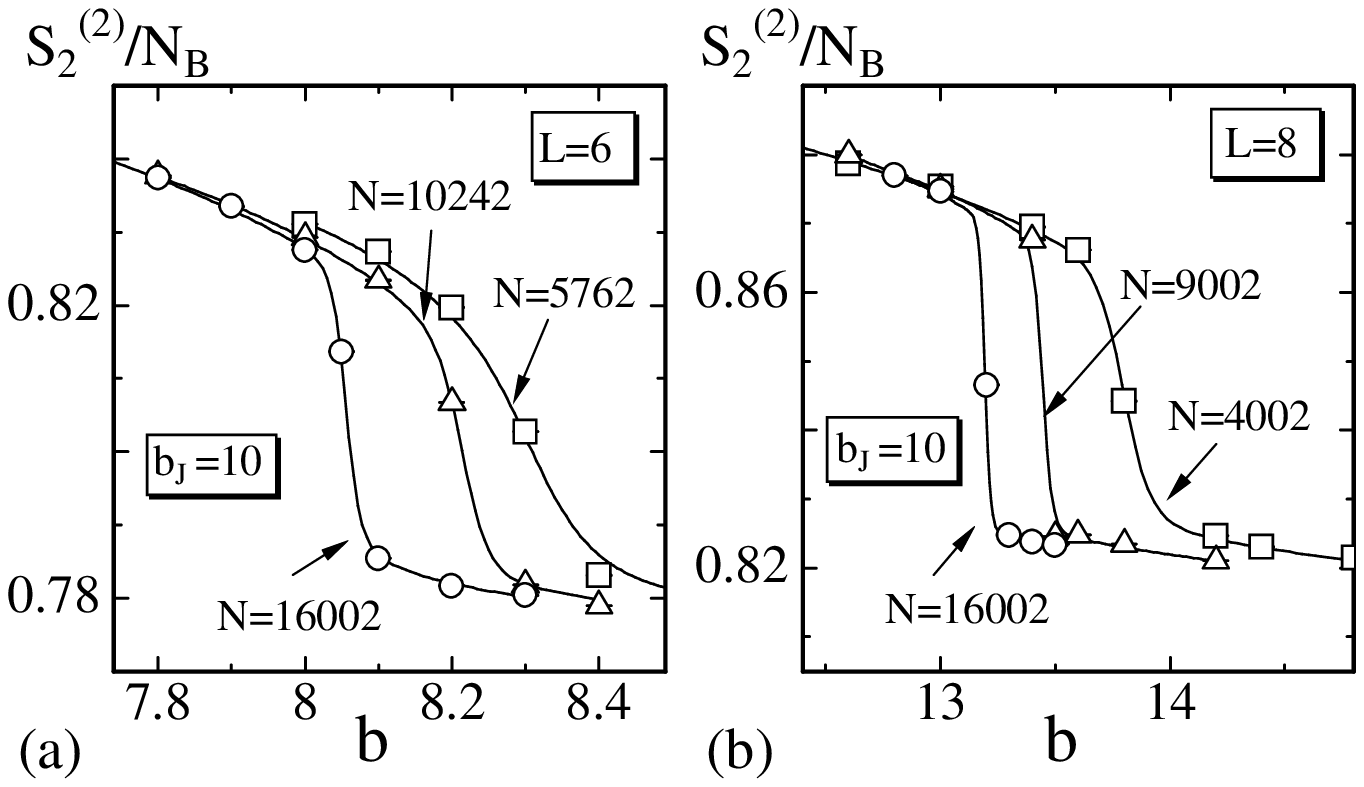}
\caption{The two-dimensional bending energy $S_2^{(2)}/N_B$ against $b$ obtained on the surfaces of (a) $L\!=\!6$ and (b) $L\!=\!8$. $S_2^{(2)}$ is defined by Eq.(\ref{two-dim-bending-energy}) and is not included in the Hamiltonian. $N_B$ is the total number of bonds where $S_2^{(2)}$ is defined.} 
\label{fig-7}
\end{figure}

On the other hand, it is possible to define a two-dimensional bending energy
\begin{equation}
\label{two-dim-bending-energy}
 S_2^{(2)}\!=\!\sum_{\langle\langle ij \rangle\rangle}(1-{\bf n}_i \cdot {\bf n}_j) 
\end{equation}
on the surface, where ${\bf n}_i$ is the unit normal vector of the triangle $i$, and $\sum_{\langle\langle ij \rangle\rangle}$ is the sum over all nearest neighbor triangles ${\langle\langle ij \rangle\rangle}$ on the surface except at the junctions. Although $S_2^{(2)}$ is not included in the Hamiltonian, the transition can be reflected in $S_2^{(2)}$. Figures \ref{fig-7}(a) and \ref{fig-7}(b) show $S_2^{(2)}/N_B$ against $b$ obtained on the surfaces of $L\!=\!6$ and $L\!=\!8$, where $N_B$ is the total number of bonds where $S_2^{(2)}$ is defined. Note that the bonds where $S_2^{(2)}$ is undefined are those where $S_J$ is defined. We immediately find the expected behavior of $S_2^{(2)}$, which again indicates a discontinuous transition.   

\begin{figure}[htb]
\centering
\includegraphics[width=10cm]{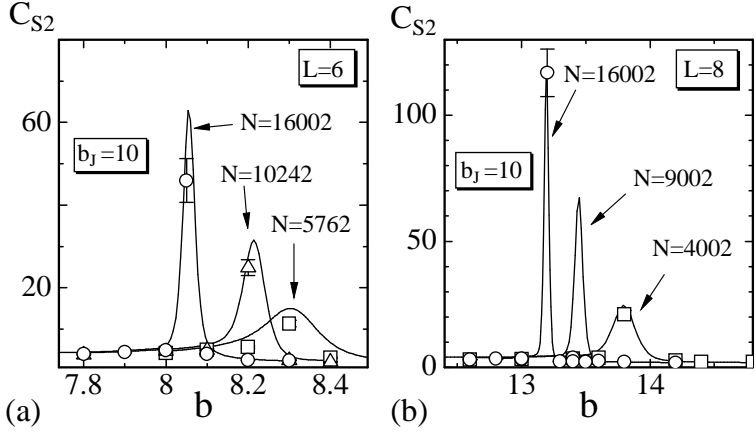}
\caption{The specific heat $C_{S_2}$ for $S_2$ against $b$ obtained on the surfaces of (a) $L\!=\!6$ and (b) $L\!=\!8$. $C_{S_2}$ is defined by Eq.(\ref{specific-heat-1}). The error bars are the statistical error, which is obtained by the binning analysis.} 
\label{fig-8}
\end{figure}

The specific heat for $S_2$ defined by 
\begin{equation}
\label{specific-heat-1}
C_{S_2} \!=\! {b^2\over N_S^\prime} \langle \; \left( S_2 \!-\! \langle S_2 \rangle\right)^2\rangle
\end{equation}
has an anomalous behavior if the model has the phase transition. Figures \ref{fig-8}(a) and \ref{fig-8}(b) show $C_{S_2}$ against $b$ obtained on the surfaces of $L\!=\!6$ and $L\!=\!8$. Sharp peaks of $C_{S_2}$ are considered to be a sign of the phase transition. 

\begin{figure}[htb]
\centering
\includegraphics[width=10cm]{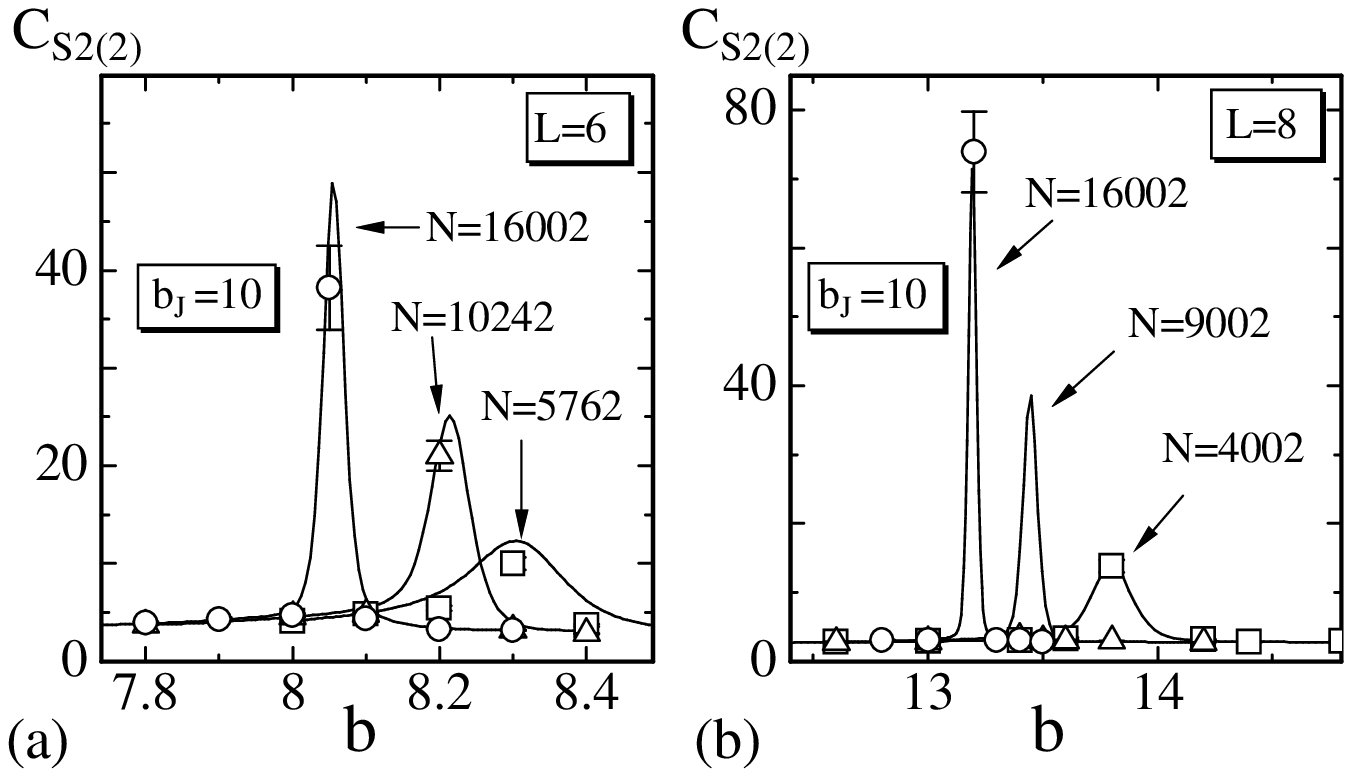}
\caption{The specific heat $C_{S_2^{(2)}}$ for $S_2^{(2)}$ against $b$ obtained on the surfaces of (a) $L\!=\!6$ and (b) $L\!=\!8$. $C_{S_2^{(2)}}$ is defined by Eq.(\ref{specific-heat-2}). The error bars are the statistical error, which is obtained also by the binning analysis.} 
\label{fig-9}
\end{figure}

Another specific heat, which is the variance of $S_2^{(2)}$, can also be defined such that
\begin{equation}
\label{specific-heat-2}
 C_{S_2^{(2)}} \!=\! {1 \over N^\prime} \langle \; ( S_2^{(2)} \!-\! \langle S_2^{(2)} \rangle)^2 \rangle, 
\end{equation}
where $N^\prime\!=\!N\!-\!6N_J\!+\!12$ is the total number of vertices $N$ minus $6N_J\!-\!12$ the number of vertices nearest to the junctions. Figures \ref{fig-9}(a) and \ref{fig-9}(b) show $C_{S_2^{(2)}}$ against $b$ obtained on the surfaces of $L\!=\!6$ and $L\!=\!8$. We find that  $C_{S_2^{(2)}}$ has the same anomalous behavior as that of $C_{S_2}$ in Figs. \ref{fig-8}(a) and \ref{fig-8}(b).

\begin{figure}[htb]
\centering
\includegraphics[width=10cm]{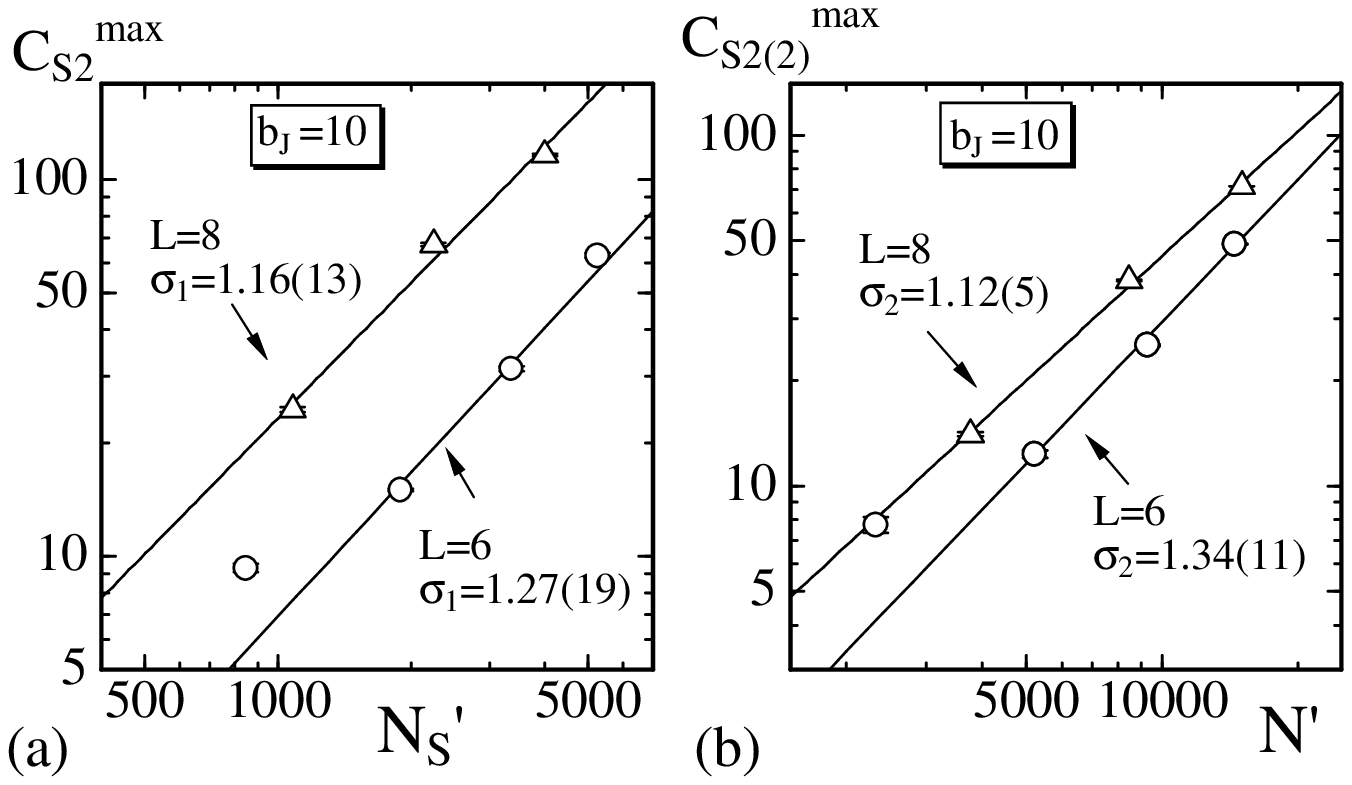}
\caption{ Log-log plots of (a) $C_{S_2}^{\rm max}$ against $N_S^\prime$ and (b) $C_{S_2^{(2)}}^{\rm max}$ against $N^\prime$ obtained on the surfaces of $L\!=\!6$ and $L\!=\!8$. The straight lines are drawn by fitting the largest three data of $C_{S_2}^{\rm max}$ and $C_{S_2^{(2)}}^{\rm max}$ to Eq.(\ref{scaling-exponents}). The peak values and the statistical errors were obtained by multihistogram reweighting.} 
\label{fig-10}
\end{figure}

In order to see the scaling property of $C_{S_2}$ and $C_{S_2^{(2)}}$, we plot the peak values  $C_{S_2}^{\rm max}$ against $N_S^\prime$ and $C_{S_2^{(2)}}^{\rm max}$ against $N^\prime$ in Figs. \ref{fig-10}(a) and \ref{fig-10}(b), respectively, in a log-log scale. The error bars on the peak values are the statistical errors, which were obtained by the so-called binning analysis. The straight lines were drawn by fitting the three largest data values to
\begin{equation}
\label{scaling-exponents}
C_{S_2}^{\rm max} \propto \left( N_S^{\prime}\right)^{\sigma_1}, \quad C_{S_2^{(2)}}^{\rm max} \propto \left( N^\prime\right)^{\sigma_2}, \quad 
\end{equation}
where $\sigma_1$, $\sigma_2$ are critical exponents. Thus, we have
\begin{eqnarray}
\label{exponents-values}
\sigma_1=1.27\pm 0.19, \quad \sigma_2=1.34\pm 0.11, \quad (L=6), \nonumber \\
\sigma_1=1.16\pm 0.13, \quad \sigma_2=1.12\pm 0.05, \quad (L=8). 
\end{eqnarray}
These values are slightly larger than $\sigma\!=\!1$, however, they are consistent with the discontinuous transition, which was already expected by all of the above-presented results. 

\section{Summary and Conclusion}\label{Conclusion}
We have investigated phase transitions between the smooth phase and the crumpled phase of a surface model with elastic chains joined to each other at junctions by using the canonical Monte Carlo simulation technique. The model is defined on uniformly triangulated spheres, which were obtained by splitting the triangles of the icosahedron. A compartmentalized structure was built on the surface, and the boundary of the compartment forms the skeletons, which have a one-dimensional bending energy and maintain shape of the surface. No two-dimensional bending energy is assumed on the surface except at the junctions. We assumed a high elasticity at the junctions; $b_J\!=\!10$, and that the length $L$ of chains between junctions is relatively small; $L\!=\!6$ and  $L\!=\!8$. 

A first-order transition has been found. The bending energy $S_2$ of the chains and the two-dimensional bending energy $S_2^{(2)}$, which is not included in the Hamiltonian, are found to have a jump at finite $b$. The corresponding specific heats have an anomalous peak, which is typical of a first-order transition. The mean square size $X^2$ also has a clear jump at the transition point, indicating that the transition distinguishes the swollen (or smooth) phase from the collapsed (or crumpled) phase. 

The fact that a skeleton surface model has a phase transition is remarkable because it is well-known  that we can see no phase transition in one-dimensional objects which are governed by a Hamiltonian for the local interactions. The elasticity of the junctions are considered to be a reason for the phase transition. The chains in our model share the two-dimensional Gaussian bond potential $S_1$ and the two-dimensional bending energy $S_J$ at the boundaries, which are the junctions. However, the two-dimensional $S_1$ seems to play no significant role in the transition, because a skeleton surface model with a one-dimensional bond potential also undergoes a phase transition \cite{KOIB-ICIC-2006}. It is expected that the skeleton surface model has a phase transition even without $S_1$ as in HPK model \cite{KOIB-NPB-2006}.     

The phase structure of the skeleton surface model seems to be significantly dependent on the junctions. Therefore, a skeleton surface model with rigid junctions is interesting. A rigid junction is a junction that has not only the infinite bending elasticity but also the infinite in-plane elasticity and, consequently, it is different from the junctions in this paper, which have finite in-plane elasticity even in the limit of $b_J\!\to\! \infty$. The fluid nature can also be incorporated in the skeleton surface model with the dynamical triangulation MC technique, and the phase structure of the model seems to be changed. Many interesting points remain to be studied in skeleton surface models.

\vspace*{3mm}
\noindent
{\bf Acknowledgments}\\
This work is supported in part by a Grant-in-Aid for Scientific Research, No. 15560160.  



\end{document}